%% file: main.tex
\documentclass[conference,a4paper]{IEEEtran}
\IEEEoverridecommandlockouts
\usepackage{amsmath,amssymb,amsfonts}
\usepackage{algorithmic}
\usepackage{graphicx}
\graphicspath{{./graphics/}}
\usepackage{textcomp}
\usepackage{xcolor}
\usepackage[colorlinks=true, urlcolor=blue, citecolor=blue, linkcolor=darkgray]{hyperref}

\usepackage{orcidlink}

\usepackage[exponent-product = \cdot]{siunitx}
\input{acronyms}

\usepackage[style=numeric-comp, 
            sorting=none,
            maxnames=1,
            minnames=1,
            maxbibnames=3,
            backend=biber,
            url=false,
            uniquelist=false,
            uniquename=false,
            doi=true,        
            url=false,       
            eprint=false,    
            isbn=false,      
            ]{biblatex} 

\DeclareSourcemap{
  \maps[datatype=bibtex]{
    \map{
      \step[fieldset=note, null]       
      \step[fieldset=eprint, null]     
      \step[fieldset=url, null]        
    }
  }
}

\def\BibTeX{{\rm B\kern-.05em{\sc i\kern-.025em b}\kern-.08em
    T\kern-.1667em\lower.7ex\hbox{E}\kern-.125emX}}
\begin{document}

\title{Gradient-Field Frequency Encoding for 1D Localization of Magneto-Mechanical Resonators
\thanks{This project is supported by the Deutsche Forschungsgemeinschaft (DFG, German Research Foundation) - SFB 1615 - 503850735.}
}

\author{

\IEEEauthorblockN{
Fabian Mohn\,\orcidlink{0000-0002-9151-9929}\IEEEauthorrefmark{1}\IEEEauthorrefmark{2},
Martin M\"oddel\,\orcidlink{0000-0002-4737-7863}\IEEEauthorrefmark{1}\IEEEauthorrefmark{2}, 
and Tobias Knopp\,\orcidlink{0000-0002-1589-8517}\IEEEauthorrefmark{1}\IEEEauthorrefmark{2}\IEEEauthorrefmark{3}\\
\small{E-mail: fabian.mohn@tuhh.de}
\vspace*{1mm}
}

\IEEEauthorblockA{\IEEEauthorrefmark{1}Institute for Biomedical Imaging, Hamburg University of Technology, Hamburg, Germany\\
\IEEEauthorrefmark{2}Section for Biomedical Imaging, University Medical Center Hamburg-Eppendorf, Hamburg, Germany\\
\IEEEauthorrefmark{3}Fraunhofer Research Institution for Individualized Medical Technology and Engineering IMTE, Lübeck, Germany}
\vspace*{-5mm}
}

\maketitle

\begin{abstract}
\Acp{MMR} have emerged as promising passive sensors for wireless localization and remote inductive sensing in industrial and medical environments. However, fundamental limits on achievable localization accuracy using the spatial sensitivity profiles of receive coils remain a key challenge. In this work, we demonstrate that an \ac{MMR} -- comprising two permanent magnets in an anti-parallel arrangement -- exhibits a measurable shift in its mechanical resonance frequency in response to an externally applied magnetic field, a property we exploit for spatial encoding. We derive and experimentally validate a method that maps frequency shifts of 1D positions within a known magnetic gradient field, demonstrating accurate localization along a linear trajectory. This gradient-field frequency encoding approach constitutes a conceptually distinct alternative to conventional sensitivity encoding with receiver coil arrays, with the potential to simplify hardware requirements while improving localization accuracy.
\end{abstract}

\begin{IEEEkeywords}
magneto-mechanical resonator (MMR), passive, wireless, inductive sensing, miniature sensor, 1D field encoding
\end{IEEEkeywords}

\section{Introduction}
\label{sec:intro}
Miniature \acp{MMR} have recently been shown to be versatile sensors for applications requiring passive and remote localization (position and orientation)~\cite{gleich_miniature_2023, fischer_magneto-oscillatory_2024} and sensing of physical quantities such as temperature, pressure, and magnetic field strength~\cite{fischer_miniature_2024, merbach_wireless_2025}. These devices transduce mechanical resonance shifts into wirelessly detectable frequency signatures that are particularly attractive for constrained environments. Despite this promise, current \ac{MMR} implementations face open challenges in device miniaturization, robust signal acquisition and reliable localization under low \ac{SNR} conditions, and the computational effort of real-time tracking and sensing at high frame rates.

Conventionally, \acp{MMR} are localized by exploiting sensitivity encoding in combination with a receiver coil array~\cite{gleich_miniature_2023, fischer_miniature_2024}. In sensitivity encoding-based approaches, the spatial sensitivity profiles of individual coil elements encode position information into inductively measured signal amplitudes, enabling reconstruction of the \ac{MMR} location through an inverse problem. As spatial high-frequency features of the sensitivity profiles inherently smooth out with distance from the coil, scaling this approach towards a larger number of coil elements incurs a cost in both the technical complexity and \ac{SNR}~\cite{ohliger_ultimate_2003,wiesinger_electrodynamics_2004,keil_massively_2013}. Alternative or complementary encoding mechanisms are therefore of considerable interest to replace sensitivity encoding entirely or to augment it and improve localization robustness when coil arrays are available, following a paradigm well established in \ac{MRI}~\cite{hutchinson_fast_1988,sodickson_simultaneous_1997,pruessmann_sense_1999}.

\begin{figure}[t]
	\centering \includegraphics[width=1.0\columnwidth]{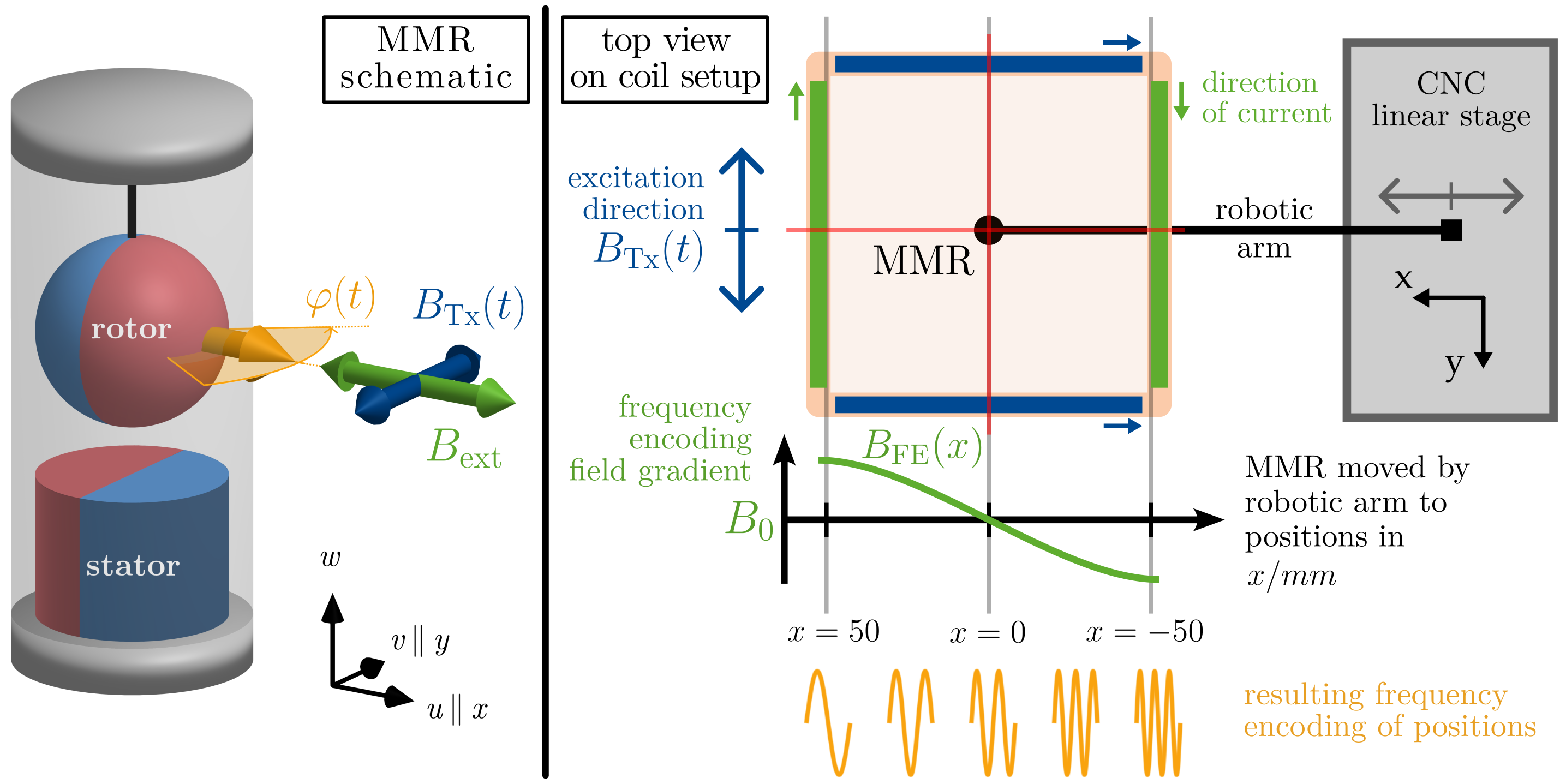}
	\caption{Experimental setup showing the MMR and a top view of the orthogonal coil system. A first coil pair generates a magnetic field gradient along the $x$-axis, whereas a homogeneous excitation field is established along the $y$-axis. The distance between the inner coils is $100$\;mm, and the MMR or magnetic field probe can be positioned along the $x$-axis using a robotic positioning system.}
	\label{fig:setup}
\end{figure}

In this work, we first demonstrate that an \ac{MMR}, comprising two small permanent magnets in an anti-parallel arrangement, exhibits a shift in its mechanical resonance frequency in response to an external magnetic field, as previously shown for the related magneto-oscillatory cantilever by Fischer~et~al.~\cite{fischer_miniature_2024}. Building on this field dependence, we propose a spatial encoding scheme that maps resonance frequency shifts to 1D positions within a known magnetic gradient field. We validate the approach experimentally along a linear trajectory, achieving a mean localization error of \SI{1.6}{\mm} with a systematic bias of \SI{0.2}{\mm} over \SI{80}{\mm} at a gradient strength of \SI{28.8 \pm 0.6}{\milli\tesla\per\meter}. This gradient-field encoding constitutes a conceptually distinct alternative to sensitivity encoding, with the potential to simplify hardware requirements while maintaining millimeter-scale localization accuracy.

\section{Methods}
\label{sec:methods}

In its original conception, an MMR consists of two magnets~\cite{gleich_miniature_2023}. One magnet (stator) is attached to the base of the MMR housing, while the other (rotor) is suspended by a thin filament and is free to oscillate around the filament axis only, as shown in \ref{fig:setup}. In its equilibrium state, the magnetic moments are aligned antiparallel in the $uv$-plane, orthogonal to the filament axis $w$. In plane deflection from the rotor's equilibrium orientation is specified by the angle $\varphi$. Assuming magnetic dipole interaction between rotor and stator, in the absence of any external magnetic field, the equation of motion of the idealized MMR is given by
\begin{equation}
    \ddot{\varphi} + \frac{\omega_\mathrm{nat}}{Q} \dot{\varphi} + {\omega_\mathrm{nat}}^2 \sin\varphi  = 0,
\end{equation}
as derived in the supplementary material of~\cite{gleich_miniature_2023}. Here, $Q$ denotes the quality factor and $\omega_\mathrm{nat}$ the natural angular frequency. A rotor displaced from equilibrium undergoes damped torsional oscillation, returning to its equilibrium orientation as the amplitude decays. Due to the nonlinear nature of the dynamical system, the instantaneous angular frequency~\cite{boashash_estimating_1992} $\omega(t)$ increases monotonically with time, converging asymptotically toward a limiting angular frequency $\omega_\mathrm{lim}$ as the oscillation amplitude decays and the system approaches its linear regime. For similar systems an analytical relation between $\omega_\mathrm{nat}$ and $\omega(t)$ has been derived~\cite{zhao_instantaneous_2018}, hence we expect changes in $\omega_\mathrm{nat}$ be directly observable in the instantaneous angular frequency traces $\omega(t)$.

The natural angular frequency is defined by $\omega_\mathrm{nat} = \sqrt{\frac{m_\mathrm{s} B}{I}}$, where $m_\mathrm{s}$ is the magnetic moment of the stator, $B$ is the $u$-component of the magnetic field at the rotor position and $I$ is the moment of inertia of the rotor~\cite{gleich_miniature_2023}. This square-root dependence on the local magnetic field has been experimentally validated by Faltinath~et~al.~\cite{faltinath_natural_2025}, who demonstrated accurate prediction of $\omega_\mathrm{nat}$ as a function of rotor-stator distance. In this work, we adopt this approach to model the dependency on the magnetic field at the rotor position with
\begin{equation}\label{eq:wnat}
    \omega_\mathrm{nat}(B) = \alpha \sqrt{B},
\end{equation}
where $\alpha$ is a lumped parameter that incorporates the remaining magnetic and mechanical properties of the MMR. In the scenario considered here, $B$ represents the superposition of the stator's own field at the position of the rotor, $B_0$, and an external magnetic field component $B_\mathrm{ext}$, such that $B = B_0 + B_\mathrm{ext}$. 
The external field $B_\mathrm{ext}$ is assumed quasi-static relative to the oscillation period of the \ac{MMR}, such that it acts as a slowly varying parameter rather than a dynamic excitation. 
The stator field $B_0$ is approximated under the magnetic dipole assumption
as $B_0 = \frac{\mu_0 m_\mathrm{s}}{4\pi d^3}$, where $\mu_0$ is the vacuum permeability, $m_\mathrm{s}$ is the magnetic moment of the stator, and $d$ is the center-to-center distance between the two magnets.

We introduce a frequency-encoding field $B_\mathrm{FE}(x) = G x$, where $G$ denotes the gradient strength along the spatial coordinate $x$. An \ac{MMR} placed at position $x$ is exposed to a unique local field $B$, giving rise to a spatially dependent instantaneous frequency $\omega(t)$. Evaluating $\omega(t)$ at a fixed time point $t_0$ across all positions yields the frequency-position relationship $\omega_{t_0}(x)$, which serves as the forward model. Position estimation is then formulated as the inverse problem: given an observed frequency $\omega(t_0)$, the corresponding position $x$ is retrieved via a data-driven non-linear interpolation of $\omega_{t_0}(x)$.

\begin{figure}[b!]
	\centering \includegraphics[width=1.0\columnwidth]{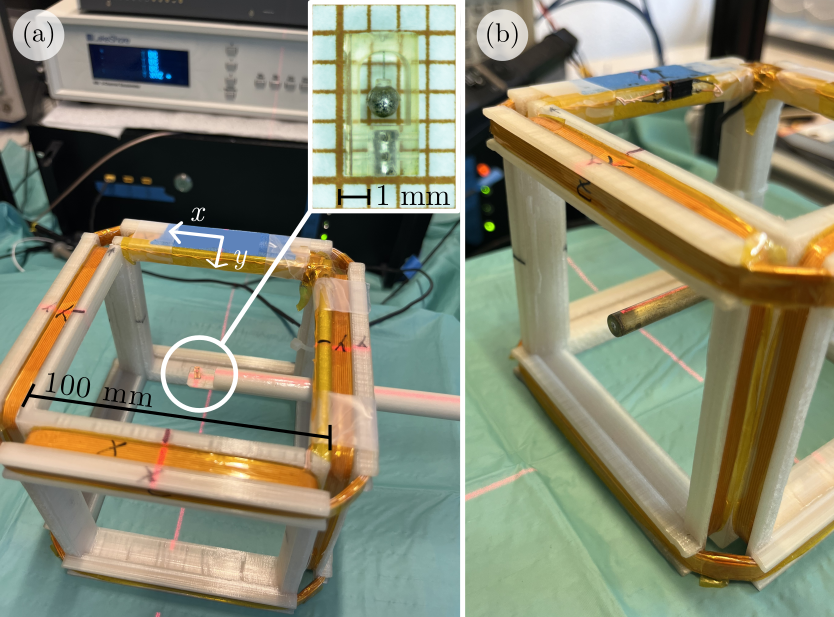}
	\caption{Images of experimental setup in 3-axis coil system. In (a), the experimental setup and an inset zoom on the MMR (on mm-scale paper) are shown. The MMR is fixed to a robotic arm for linear positioning and a cross-line laser projection is utilized for (re-)positioning after probe changes. In (b), the LakeShore 460 gaussmeter probe is shown during step-wise evaluation of the gradient field.}
	\label{fig:images}
\end{figure}

\begin{figure*}[th]
	\centering\includegraphics[width=1.0\linewidth]{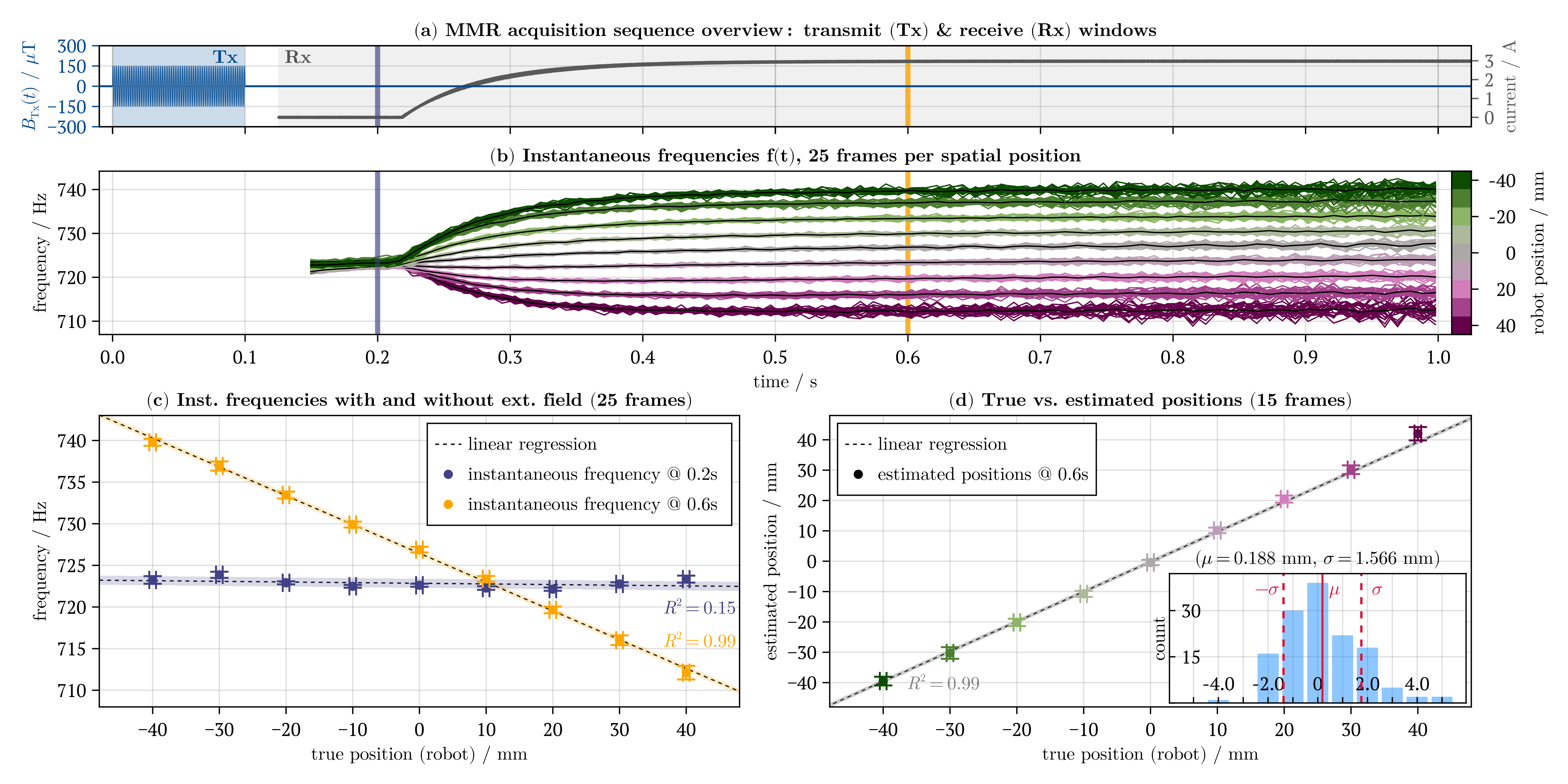}
	\caption{Spatial frequency encoding of an MMR. The sequence in (a) shows a Tx/Rx window (one frame). The current of the gradient field in $x$ is switched on at 225\;ms and builds up for about 200\;ms in this experiment. Depending on the spatial position in the gradient field, the instantaneous frequency in (b) is shifted up or down. In (c), the data at two time points (0.2\;s, blue, and 0.6\;s, orange) are shown, including a linear fit. The results in (d) show estimated positions versus the true positions of the linear positioning system, including a histogram of bias and error for the estimator.
    }
	\label{fig:results}
\end{figure*}

\subsection{Measurement Setup}
\label{sec:meas_setup}

In order to exploit the dependence of $\omega_\mathrm{nat}(x)$ most effectively, we use two orthogonal coil pairs and orientate the \ac{MMR} accordingly, as shown in \autoref{fig:setup}. The first coil pair is in Maxwell configuration, generating a gradient field $B_\mathrm{FE}(x)$ in $x$-direction. The second coil pair is in Helmholtz configuration, generating a homogeneous field $B_\mathrm{Tx}(t)$ in $y$-direction, set to \SI{150}{\micro\tesla} at \SI{725}{\Hz}. 
A small \ac{MMR} is used for all measurements, approximately \qtyproduct{5 x 2.5 x 2.5}{\mm} in size, with an estimated $f_\mathrm{nat} = \frac{\omega_\mathrm{nat}}{2\pi}$ around \SI{725}{\Hz}. The sensor consists out of two neodymium–iron–boron (NdFeB) permanent magnets, with a cylindrical stator (N50) of \SI{1.5}{\mm} height and \SI{1}{\mm} in diameter, and a spherical rotor (N52), \SI{1}{\mm} in diameter, suspended by a thin ultrahigh-molecular-weight polyethylene (PE-UHMW) type filament. An estimated value of $B_0$ is approximately \SI{25}{\milli\tesla} for these dimensions~\cite{gleich_miniature_2023}.

A cross-lines laser is used to (re-) position and align the MMR and the 3-channel gaussmeter probe (LakeShore Model 460, Lake Shore Cryotronics, Inc., USA), as shown in \autoref{fig:images}. A linear positioning system (M.500, Stepcraft GmbH \& Co. KG, Germany) drives the robotic arm to a number of $9$ equidistant $x$-positions in \SI{10}{\mm} steps from \SIrange{-40}{40}{\mm}, in identical procedure for MMR and field measurements of $B_\mathrm{FE}(x)$.

Signal acquisition is based on the RedPitaya STEMlab 125-14 system-on-chip (Red Pitaya d.o.o., Slovenia), following the experimental setup described in~\cite{faltinath_natural_2025}. The raw signals were bandpass-filtered and subsequently processed via \ac{STFT} to estimate the instantaneous frequency $f(t) = \omega(t)/(2\pi)$. At each measurement position, $25$ independent frames were acquired. The first $10$ frames were used to calibrate the inverse interpolator, while the remaining $15$ frames served as independent test data for position estimation.

\section{Results}
\label{sec:results}

Field measurements of $B_\mathrm{FE}(x)$ over all positions fall within the range \SIrange{-1.15 \pm 0.02}{1.15 \pm 0.02}{\milli\tesla}. Observations are in agreement with a linear gradient field $B_\mathrm{FE}(x)$ with a gradient strength of \SI{28.8 \pm 0.6}{\milli\tesla\per\meter}. Sources of uncertainty for position are the accuracy of the linear stage and the visual cross-laser alignment which amount to \SI{\pm 0.5}{\mm}. 
In absence of any external field, $\omega(t)$ increases over time as the instantaneous amplitude of the MMR oscillation decreases in agreement with earlier observations~\cite{faltinath_natural_2025}. Estimation of $\omega(t)$ shows increasing uncertainty as the oscillation amplitude decays.

In \autoref{fig:results}, two time points at \num{0.2} and \SI{0.6}{\s} are considered, without the presence of an external field and in the presence of the ramped-up gradient field $B_\mathrm{FE}(x)$, respectively. For the former time point, a frequency of \SI{722.8 \pm 0.6}{\Hz} over all frames and positions was observed. At $x=0$ ($B_\mathrm{ext}=0$), we observe an expected increase of $\omega(t)$ between the time points, due to the aforementioned oscillation decay. Results at the later time point show an approximately linear relation between position $x$ and $\omega(t)$.

Based on the observed frequency-position relationship, the inverse interpolator was applied to estimate the 1D position of the \ac{MMR} from the measured $\omega(t)$ (15 frames, 9 positions). A parity plot of estimated versus true position reveals a close agreement across the full range of evaluated positions, confirming the validity of the linear encoding assumption with a systematic bias of \SI{0.2}{\mm} and a standard deviation of \SI{1.6}{\mm}.

\section{Discussion}
\label{sec:conclusion}

We demonstrate gradient-field-based frequency encoding as a possible step towards an alternative to sensitivity encoding for \ac{MMR} localization. The results might establish a basis to increase accuracy for future real-time multi-dimensional tracking in industrial and medical applications. The underlying mechanism can be exploited for B-field sensing using \acp{MMR}.

A stronger $B_\mathrm{FE}$ shifts $\omega(t)$ beyond the linear regime, in which case \eqref{eq:wnat} predicts a square-root dependence~\cite{fischer_magneto-oscillatory_2024} that is accounted for by the inverse non-linear interpolator.
The position uncertainty depends mainly on the gradient strength and the uncertainty of the \ac{MMR}'s instantaneous frequency. High gradients result in lower position uncertainty. For the linear dominated regime, the gradient determines the position-frequency sensitivity of \SI{2.9}{\mm\per\Hz} and is adjustable via the applied gradient strength. 
Slow drift of $\omega(t)$ over a large number of frames due to environmental or sensing-related effects, e.g. temperature, does not compromise the proposed approach, since the encoding can be based on the \emph{relative} frequency shift induced by $B_\mathrm{FE}$, as reflected by the vertical separation of the frequency-position curves in \autoref{fig:results}(c). 

The approach currently does not require prior characterization of $B_\mathrm{FE}(x)$, but is limited to 1D localization at a fixed MMR orientation in strictly monotone fields; extension to 3D tracking will require gradient sequences similar to \ac{MRI}.

\section*{Acknowledgments}
The authors would like to thank Nora Timm for designing and fabricating the \ac{MMR} used in this study.

\clearpage

\printbibliography%

\end{document}

%% file: acronyms.tex
\usepackage[nolist,nohyperlinks]{acronym}

\begin{acronym}[ECU]

\acro{AWMPS}[AWMPS]{arbitrary waveform magnetic particle spectrometer}
\acro{ADC}[ADC]{analog-to-digital converter} 
\acro{AUC}[AUC]{area under the curve}

\acro{CT}[CT]{computed tomography}

\acro{DAC}[DAC]{digital-to-analog converter}
\acro{DFG}[DFG]{drive-field generator}
\acro{DFCs}[DFCs]{drive-field coils}
\acro{DF}[DF]{drive-field}
\acro{DA}[DA]{differential amplifier}
\acro{DLS}[DLS]{dynamic light scattering}
\acro{DTS}[DTS]{dispersion technology software}

\acro{ECD}[ECD]{equivalent circuit diagram}

\acro{FFL}[FFL]{field-free-line}
\acro{FFP}[FFP]{field-free-point}
\acro{FFR}[FFR]{field-free-region}
\acro{FFT}[FFT]{fast Fourier transform}
\acro{FOV}[FOV]{field-of-view}
\acro{FWHM}[FWHM]{full width at half maximum}

\acro{GUI}[GUI]{graphical user interface}

\acro{HCC}[HCC]{high current circuit}

\acro{ICN}[ICN]{inductive coupling network}
\acro{ISI}[ISI]{integrated signal intensity}
\acro{ICs}[ICs]{integrated circuits}
\acro{IA}[IA]{instrumentation amplifier}
\acro{ICU}[ICU]{intensive care unit} 

\acro{LC}[LC]{inductor-capacitor}
\acro{LFR}[LFR]{low-field-region}
\acro{LNA}[LNA]{low noise amplifier}

\acro{MPI}[MPI]{magnetic particle imaging}
\acro{MRI}[MRI]{magnetic resonance imaging}
\acro{MTT}[MTT]{mean-transit-time}
\acro{MNP}[MNP]{magnetic nanoparticle}
\acro{MPS}[MPS]{magnetic particle spectroscopy}
\acro{MAE}[MAE]{mean absolute error}
\acro{MSE}[MSE]{mean squared error}
\acro{MMR}[MMR]{magneto-mechanical resonator}

\acro{PSF}[PSF]{point spread function}
\acro{PNS}[PNS]{peripheral nerve stimulation}
\acro{PTT}[PTT]{pulmonary transit time}
\acro{PDI}[PDI]{polydispersity index}

\acro{Q}[Q-factor]{quality factor}

\acro{RF}[RF]{radio-frequency}
\acro{RP}[RP]{RedPitaya STEMlab 125-14}
\acro{RF-fields}[RF]{radio frequency fields}
\acro{rBV}[rBV]{relative blood-volume}
\acro{rBF}[rBF]{relative blood-flow}
\acro{rCBV}[rCBV]{relative cerebral-blood-volume}
\acro{rCBF}[rCBF]{relative cerebral-blood-flow}
\acro{RBCs}[RBCs]{red blood cells}

\acro{SNR}[SNR]{signal-to-noise ratio}
\acro{SU}[SU]{surveillance unit}
\acro{SAR}[SAR]{specific absorption rate}
\acro{SPIONs}[SPIONs]{superparamagnetic iron-oxide nanoparticles}
\acro{USPIONs}[USPIONs]{ultrasmall superparamagnetic iron-oxide nanoparticles}
\acro{SF}{selection field}
\acro{SFG}[SFG]{selection-field generator}
\acro{SEM}[SEM]{scanning electron microscope}
\acro{SSIM}[SSIM]{structural similarity index measure}
\acro{STFT}[STFT]{short-time Fourier transform}

\acro{TF}[TF]{transfer function}
\acro{THD}[THD]{total harmonic distortion}
\acro{TTP}[TTP]{time-to-peak}
\acro{TEM}[TEM]{transmission electron microscopy}
\acro{TxRx}[TxRx]{transmit-receive}

\acro{USPIO}[USPIO]{ultra-small superparamagnetic iron oxide}

\acro{VOI}[VOI]{volume of interest}
\acro{VSM}[VSM]{vibrating sample magnetometry}

\end{acronym}